\documentclass[epj,referee]{svjour}
\usepackage{graphics}
\usepackage{amssymb}

\begin{document}
\title{Disturbing synchronization: Propagation of perturbations
in networks of coupled oscillators}
\author{Dami\'an H. Zanette}
\institute{Consejo Nacional de Investigaciones Cient\'{\i}ficas y
T\'ecnicas \\ Centro At\'omico Bariloche and Instituto Balseiro,
8400 Bariloche, Argentina}
\date{Received: date / Revised version: date}
\abstract{We study the response of an ensemble of synchronized phase
oscillators to an external harmonic perturbation applied to one of
the oscillators. Our main goal is to relate the propagation of the
perturbation signal to the structure of the interaction network
underlying the ensemble. The overall response of the system is
resonant, exhibiting a maximum when the perturbation frequency
coincides with the natural frequency of the phase oscillators. The
individual response, on the other hand, can strongly depend on the
distance to the place where the perturbation is applied. For small
distances on a random network, the system behaves as a linear
dissipative medium: the perturbation propagates at constant speed,
while its amplitude decreases exponentially with the distance. For
larger distances, the response saturates to an almost constant
level. These different regimes can be analytically explained in
terms of the length distribution of the paths that propagate the
perturbation signal. We study the extension of these results to
other interaction patterns, and show that essentially the same
phenomena are observed in networks of chaotic oscillators. \PACS{
{05.45.Xt}{Synchronization; coupled oscillators}\and
{05.65.+b}{Self-organized systems}}} \maketitle

\section{Introduction}
\label{intro}

Synchronization is a paradigmatic mode of emergent collective
behaviour in ensembles of interacting dynamical elements
\cite{Pik,nos}. It arises in a broad class of real systems,
comprising from mechanical and physico-chemical processes
\cite{ph1,ph2,ph3} to biological phenomena \cite{b1,b2,b3}, and is
reproduced by a variety of mathematical models. Roughly speaking, it
consists of some kind of coherent evolution where the motions of
individual elements are correlated in time. Depending on the nature
of the individual dynamical laws and on the interactions, different
forms of synchronized states are possible. They range from full
synchronization, where all the elements follow the same orbit in
phase space, to weakly correlated forms where the ensemble splits
into almost independent clusters of mutually synchronized elements,
or where coherence manifests itself in just a few state variables or
in time averages of suitably chosen quantities \cite{kura,nos}. Full
synchronization is typical for globally coupled ensembles of
identical elements, where any two elements interact with the same
strength. When coupling is strong enough and represents an
attractive interaction, the asymptotic state where all elements share
the same orbit is stable \cite{bocca}. Under certain conditions,
stability of the fully synchronized state can also be insured for
more complex interaction patterns, where each pair of elements may or
may not interact \cite{str}. Weaker forms of synchronization are
characteristic of ensembles of non-identical dynamical elements.

A simple but quite useful model for an ensemble of coupled dynamical
elements is given by a set of $N$ phase oscillators, whose
individual dynamics in the absence of coupling is governed by $\dot
\phi =\omega$. The phase $\phi (t) \in [0,2\pi)$ rotates with
constant frequency $\omega$. This elementary representation of
periodic motion, originally introduced as a model for biological
oscillations \cite{b2}, approximates any cyclic dynamics, even in
the presence of weak coupling \cite{kura}. As for the interaction
pattern, it can be thought of as a graph, or network, with one
oscillator at each node. The graph is characterized by its adjacency
matrix ${\cal J} = \{ J_{ij} \}$. If oscillator $i$ is coupled to
oscillator $j$, i. e. if the phase $\phi_j(t)$ enters the equation
of motion of $\phi_i(t)$, we have $J_{ij}=1$, and $J_{ij}=0$
otherwise. The adjacency matrix is not necessarily symmetric and,
thus, coupling is not always bidirectional. In other words, the
interaction network is generally a directed graph. The coupled
oscillator ensemble is governed by the equations \cite{d1}
\begin{equation}
\dot \phi_i =\omega_i +k \sum_{j=1}^N J_{ij} \sin (\phi_j-\phi_i)
\end{equation}
for $i=1,\dots,N$, where $k$ is the coupling strength. The case of
global coupling, $J_{ij}=1$ for all $i$ and $j$, has been studied in
the thermodynamical limit by Kuramoto \cite{kura}, who found that,
as the coupling strength grows, the system undergoes a transition to
a state of frequency synchronization, as first predicted by Winfree
\cite{win}. The transition parameters are determined by the
distribution of natural frequencies $\omega_i$.

For identical oscillators, $\omega_i=\omega$ for all $i$, and for
all $k>0$,  the long-time asymptotic state of a globally coupled
ensemble is full synchronization. More generally, it is possible to
show that full synchronization is stable when the interaction
network is regular, i. e. when all oscillators are coupled to
exactly the same number $z$ of neighbours \cite{str}. In this
situation, $\sum_{j} J_{ij}=z$ for all $i$.

A fully synchronized oscillator ensemble can be thought of as an
active medium in a rest-like state. Microscopically, this stable
state is sustained by the highly coherent collective dynamics of the
interacting oscillators. A key feature characterizing the dynamical
properties of the medium is determined by its response to an
external perturbation. How is the synchronized state altered as the
perturbation signal propagates through the ensemble? Which
propagation properties does coupling between oscillators establish
in the medium? The effect of external forces on ensembles of
interacting dynamical systems has been studied in detail for global
coupling, both for periodic oscillators and chaotic elements
\cite{saka,kaneko,naka,veser}. Ordered oscillator arrays have also
been considered \cite{coullet}. The above questions, however, are
especially significant for more complex interaction patterns --in
particular, for random interaction networks-- where the non-trivial
geometric structure is expected to play a relevant role in the
propagation process. Quite surprisingly, the problem seems to have
been addressed for the first time only recently \cite{kori,EPL}.

In this paper, we present numerical calculations and analytical
results for the propagation of a perturbation in an ensemble of
identical phase oscillators, governed by the equations
\begin{equation}
\label{model}
\dot \phi_i =\sum_{j=1}^N J_{ij} \sin (\phi_j-\phi_i)+
a \delta_{i1} \sin(\Omega t -\phi_i) .
\end{equation}
Without generality loss, the natural frequency of oscillators and
the coupling strength are fixed to $\omega=0$  and $k=1$,
respectively. The external perturbation is represented by an
additional oscillator of constant frequency $\Omega$, to which a
single oscillator in the ensemble --i. e., oscillator $1$-- is
coupled with strength $a$. To take advantage of certain analytical
results regarding regular graphs, we take a connection network where
each oscillator is coupled to exactly $z$ neighbours. Our main goal
is to relate the response of the ensemble to the metric properties
of the interaction network. In particular, we pay attention to the
dependence on the distance to the node where the perturbation is
applied, and on the perturbation frequency.

The paper is organized as follows. In the next section, we present
numerical results for random interaction networks of phase
oscillators. We show that the system response is, essentially, a
resonance phenomenon, and identify two regimes in the dependence on
the distance. In Sect.~\ref{sec:anal}, we reproduce the numerical
results through an analytical approach in the limit of
small-amplitude perturbations. We propose an approximation to obtain
explicit expressions which clarify the role of the network structure
in the propagation process. In Sect.~\ref{sec:ext}, we extend our
results to other geometries, including ordered networks with uni- and
bidirectional interactions and hierarchical structures. Moreover, we
show that the same propagation properties are observed in networks of
chaotic oscillators, which broadly generalizes our conclusions.
Results are summarized and commented in the final section.

\section{Numerical results}
\label{sec:num}

We have solved Eqs.~(\ref{model}) numerically, for an ensemble of
$N=10^3$ oscillators. We have considered a random regular network,
with $z=2$. The network was constructed by chosing at random the $z$
neighbours of each node. Multiple directed links between any two
nodes were avoided, and realizations which produced disconnected
networks were discarded. Most of the results presented here
correspond to a perturbation of amplitude $a=10^{-3}$ and various
frequencies, typically ranging from $\Omega \sim 10^{-2}$ to $10$.
The integration $\Delta t$ in our numerical algorithm was chosen such
that $\Omega^{-1} \gg \Delta t$.

The ensemble was prepared in the state of full synchronization, with
$\phi_i = \phi_j$ for all $i$ and $j$. Before recording the
evolution of the phases $\phi_i$, an interval much longer than
$\Omega^{-1}$ was left to elapse. After this interval, any transient
behaviour due to the combined effect of the dissipative mechanisms
inherent to the coupled oscillator dynamics and the external
perturbation had relaxed and the system had reached a regime of
steady evolution. Numerical results show that, in this regime, each
phase $\phi_i(t)$ oscillates around the average $\bar \phi (t) =
N^{-1} \sum_i \phi_i (t)$, seemingly with harmonic motion of
frequency $\Omega$. Our aim was to quantitatively characterize the
departure from full synchronization due to the response to the
external perturbation. As a measure of this departure for each
individual oscillator $i$, we considered the time-averaged mean
square deviation in $\phi$-space, defined as
\begin{equation} \label{sigma}
\sigma_{\phi_i} =  \left[\left\langle (\phi_i - \bar\phi)^2
\right\rangle \right]^{1/2},
\end{equation}
where $\langle \cdot \rangle$ denotes time averages over
sufficiently long intervals. The mean square deviation
$\sigma_{\phi_i}$ is a direct measure of the amplitude of the motion
of each phase with respect to the average $\bar\phi$. In the fully
synchronized state, $\sigma_{\phi_i}=0$ for all $i$.

It turns out that, for a given realization of the interaction network
and a fixed value of the frequency $\Omega$, $\sigma_{\phi_i}$ has a
rather well defined dependence on the distance $d_i$ from oscillator
$1$, where the external perturbation is applied, to oscillator $i$.
The distance $d_i$ is defined as the number of links along the
shortest directed path starting at oscillator $1$ and ending at $i$.
On the other hand, especially for large distances $d_i$,
$\sigma_{\phi_i}$ may strongly depend on the specific realization of
the interaction network. In view of the well defined dependence of
the mean square deviation on the distance, and for clarity in the
notation, from now on we drop the index $i$ which identifies
individual oscillators.

\begin{figure}
\resizebox{\columnwidth}{!}{\includegraphics*{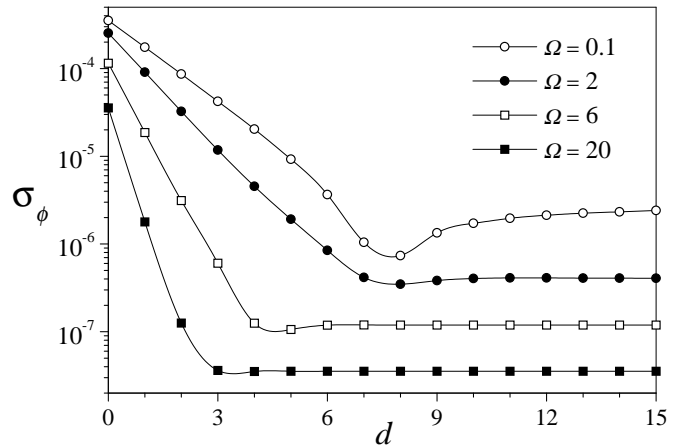}} \caption{Mean
square deviation from full synchronization as a function of the
distance from the oscillator where the external perturbation is
applied, for various values of the frequency $\Omega$. As a guide to
the eye, spline interpolations are shown as curves.} \label{f1}
\end{figure}

Figure \ref{f1} shows results for the mean square deviation
$\sigma_{\phi}$ as a function of the distance $d$ for several
frequencies $\Omega$ and a fixed interaction network. In this
realization of the network, the maximal distance between oscillator
$1$ and any other oscillator is $d_{\rm max}=15$. For each value of
$d$, the individual values of $\sigma_{\phi}$ have been averaged
over all the oscillators at that distance from oscillator $1$. It is
apparent that the mean square deviation exhibits two well
differentiated regimes as a function of the distance. For small $d$,
$\sigma_{\phi}$ decreases exponentially, at a rate that sensibly
depends on the frequency $\Omega$. In this regime, the perturbation
is increasingly damped as it propagated through the system. At large
distances, on the other hand, $\sigma_{\phi}$ is practically
independent of $d$. It is this large-distance value of
$\sigma_{\phi}$ which, typically, shows considerable variations
between different realizations of the interaction network. The
transition between the two regimes is mediated by a zone where
$\sigma_{\phi}$ attains a minimum, which is sharper for smaller
frequencies. Thus, the mean square deviation varies
non-monotonically with the distance.

\begin{figure}
\resizebox{\columnwidth}{!}{\includegraphics*{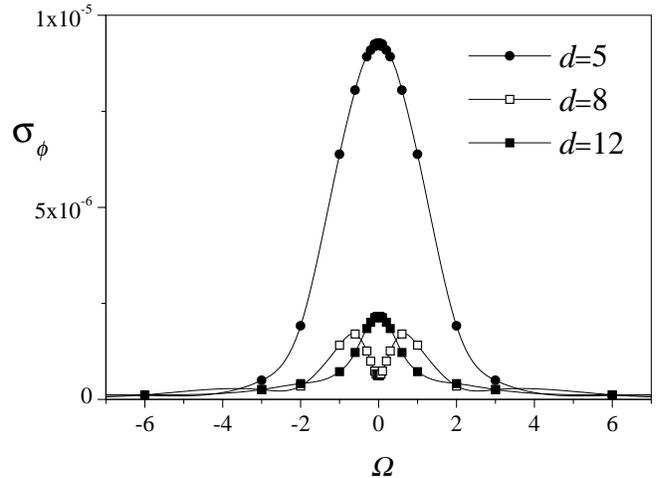} } \caption{Mean
square deviation from full synchronization as a function of the
frequency of the external perturbation, at various distances from the
node at which the perturbation is applied. Curves, added for clarity,
are spline interpolations.} \label{f2}
\end{figure}

The dependence of $\sigma_{\phi}$ on the frequency $\Omega$ for fixed
distance reveals that the response of the oscillator ensemble to the
external perturbation is, essentially, a resonance phenomenon. In
Fig.~\ref{f2} we plot $\sigma_{\phi}$ as a function of $\Omega$ for
three values of $d$. For small and large distances, the mean square
deviation from full synchronization displays a symmetric peak around
the natural frequency of the individual oscillators ($\omega=0$).
This shows that the response of the ensemble is maximal when the
external perturbation varies with the same frequency of the
elementary components of the system. For intermediate distances,
however, an anomaly appears. While for large frequencies the
response decreases as expected, the resonance peak is replaced by a
local minimum at $\Omega=\omega$. The response is now maximal at two
symmetric values around the natural frequency of the oscillators.
This anomaly is directly related to the fact that the minimum in
$\sigma_{\phi}$ for intermediate distances (Fig.~\ref{f1}) is sharp
for small $\Omega$ and becomes much less distinct as the frequency
of the external perturbation increases. This effect, combined with
the overall decrease of $\sigma_{\phi}$ as $\Omega$ grows, implies a
non-monotonic dependence $\Omega$ at such distances, which results
into the appearance of the double peak.

\begin{figure}
\resizebox{\columnwidth}{!}{\includegraphics*{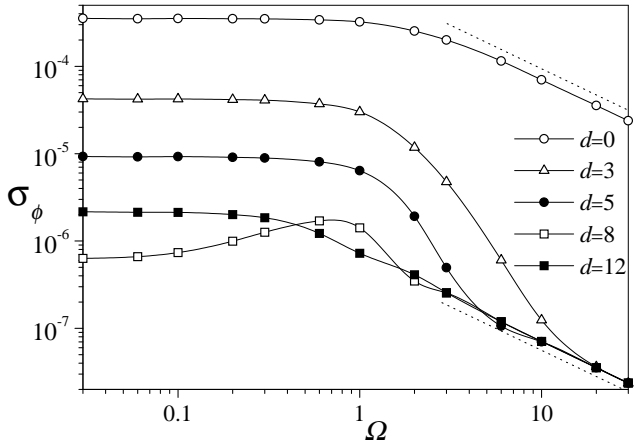} } \caption{The
same as in Fig.~\ref{f2}, in log-log scale. Data for $d=0$ and $d=3$
are also shown. Dotted lines have slope $-1$.} \label{f3}
\end{figure}

The double-logarithmic plot of Fig.~\ref{f3} reveals the
large-$\Omega$ behaviour of the mean square deviation from full
synchronization. Beyond the resonance zone, $\sigma_{\phi}$
decreases with the frequency as $\Omega^{-1}$ for all distances.
Note, however, that the value of $\sigma_{\phi}$ in the
large-$\Omega$ regime is the same for all $d>0$, while for $d=0$
--i. e., at oscillator $1$, where the external perturbation is
applied-- $\sigma_{\phi}$ is three orders of magnitude larger.

\begin{figure}
\resizebox{\columnwidth}{!}{\includegraphics*{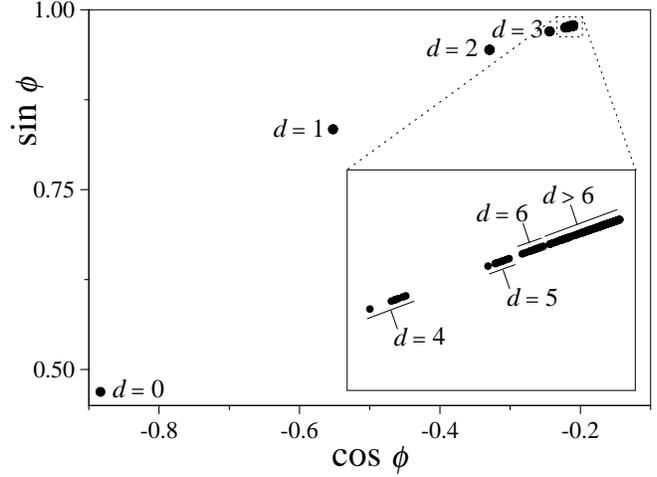} }
\caption{Snapshot of individual phases on the plane $(\cos \phi ,
\sin \phi)$. Labels indicate the distance to the oscillator where
the external perturbation is applied. The insert shows a close-up of
oscillators with $d>3$, revealing the fine cluster structure for
larger distances.} \label{f4}
\end{figure}

The fact that the mean square deviation $\sigma_{\phi}$ shows a well
defined dependence on the distance does not necessarily imply that
the individual motions of oscillators with the same value of $d$ are
related in any specific way. As a matter of fact, being a time
average, $\sigma_{\phi}$ bears no information about possible
correlations between the instantaneous state of different
oscillators. To detect such correlations we have inspected
successive snapshots of the set of individual phases plotted on the
unit circle, i. e. on the plane $(\cos \phi, \sin \phi)$. Figure
\ref{f4} shows one of these snapshots for the same system of
Figs.~\ref{f1} to \ref{f3}. It turns out that, actually, there is a
strong correlation between the positions of oscillators with the
same value of $d$. According to their distance, they form clusters
of gradually decreasing deviations and growing dispersion. Clusters
with $d<4$ are so compact that they cannot be resolved into single
elements in the scale of the main plot of Fig.~\ref{f4}. For larger
distances, clusters are relatively more disperse, as shown in the
insert. Clustering in the distribution of phases reveals that, for a
given value of $d$, oscillations around the average phase $\bar \phi$
occur coherently. In the next section we show analytically that, in
fact, these oscillations are very approximately in-phase.

\begin{figure}
\resizebox{\columnwidth}{!}{\includegraphics*{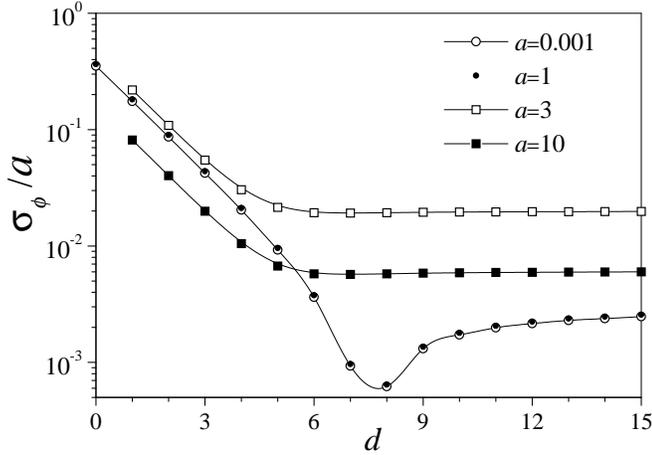} }
\caption{Normalized mean square deviation, $\sigma_{\phi} / a$, for
various values of the perturbation amplitude $a$, as a function of
the distance to the oscillator where the external perturbation is
applied. The perturbation frequency is $\Omega=0.1$. } \label{f5}
\end{figure}

Finally, we have studied the dependence of the response of the system
on the perturbation amplitude $a$. Over a wide range, the mean square
deviation of individual oscillators results to be proportional to
the amplitude, $\sigma_{\phi} \propto a$. As an illustration,
Fig.~\ref{f5} shows the ratio $\sigma_{\phi} /a$ as a function of
the distance for a perturbation frequency $\Omega=0.1$. Curves
collapse for all $a \lesssim 2$. Only for $a>2$ do we find
significant departures from the small-amplitude regime. For $a=3$ the
response of the whole system is relatively increased, especially, for
intermediate and large distances. Finally, for $a=10$ we have an
overall saturation of the response, and the ratio $\sigma_{\phi} /a$
decreases.

In the next section, we show that most of the numerical results
presented here can be analytically explained by means of a
small-amplitude approximation of Eq.~(\ref{model}). Our analytical
approach reveals the role of the interaction structure in the
response of the system to the external perturbation.

\section{Analytical results}
\label{sec:anal}

\subsection{Small-amplitude limit}
\label{sec:anal1}

The numerical results presented in Fig.~\ref{f5} suggest that the
limit of small external perturbation ($a\ll 1$) bears significant
information on a wide range of values for the amplitude $a$. It is
therefore worthwhile to study this limit analytically, representing
the instantaneous state of the oscillator ensemble as a perturbation
of order $a$ to full synchronization. Introducing, as in the
previous section, the average phase $\bar \phi (t) = N^{-1} \sum_i
\phi_i (t)$, we write the phase of oscillator $i$ as a perturbation
of order $a$ to $\bar \phi$,
\begin{equation} \label{phii}
\phi_i (t)=\bar \phi (t)+a \psi_i (t),
\end{equation}
with $\sum_i \psi_i =0$. In turn, the average $\bar \phi$ is
expected to vary around a constant phase $\phi_0$, with fluctuations
of amplitude proportional to $a$:
\begin{equation} \label{barphi}
\bar \phi (t)= \phi_0 +a \Phi (t).
\end{equation}
Without generality loss, we take $\phi_0=0$.

Replacing Eqs.~(\ref{phii}) and (\ref{barphi}) in (\ref{model}), and
expanding to the first order in the perturbation amplitude $a$, we
get
\begin{equation} \label{Phi}
\dot \Phi =\frac{1}{N} \sum_{ij} J_{ij} (\psi_j-\psi_i)+\frac{1}{N}
\exp({\rm i}\Omega t)
\end{equation}
for the average phase deviation, and
\begin{equation} \label{psi}
\dot \psi_i= - \dot \Phi +\sum_j J_{ij} (\psi_j-\psi_i) + \delta_{i1}
\exp({\rm i}\Omega t)
\end{equation}
for the individual deviations. For simplicity in the mathematical
treatment, we have replaced $\sin (\Omega t)$ by $\exp({\rm i}\Omega
t)$. Focusing on the case where each oscillator is coupled to
exactly $z$ neighbours, the above equations can be simplified taking
into account that $\sum_j J_{ij} (\psi_j-\psi_i) =-z \psi_i+ \sum_j
J_{ij} \psi_j$ and $\sum_{ij} J_{ij} (\psi_j-\psi_i) = \sum_{ij}
J_{ij} \psi_j$.

Note that the time derivative of the average phase deviation $\Phi$
enters the equation of motion for the individual deviations $\psi_i$,
Eq.~(\ref{psi}), as a kind of external force acting homogeneously
over the whole ensemble. According to Eq.~(\ref{Phi}), this
effective force is of order $N^{-1}$. As we show later, it dominates
the response of oscillators at large distances, where the effect of
the perturbation signal propagated through the network is lower.

Equation (\ref{psi}) admits solutions of the form $\psi_i(t)= A_i
\exp({\rm i} \Omega t)$, corresponding to steady harmonic motion at
the frequency of the external perturbation. These steady solutions
are expected to represent the motion once transients have elapsed.
The complex amplitudes $A_i$ satisfy a set of linear equations,
which can be cast in matrix form as
\begin{equation} \label{matrix}
{\cal L} {\bf A} = {\bf b},
\end{equation}
with ${\bf A}= (A_1,A_2,\dots ,A_N)$. The elements of the matrix
$\cal L$ and of the vector $\bf b$ are, respectively,
\begin{equation} \label{Lij}
L_{ij}=(z+{\rm i}\Omega) \delta_{ij}-J_{ij}+\frac{1}{N}\sum_k J_{kj}
\end{equation}
and
\begin{equation}
b_i=\delta_{i1}-\frac{1}{N}.
\end{equation}
For a given interaction network, the amplitudes $ {\bf A} = {\cal
L}^{-1}{\bf b}$ can be found, for instance, numerically. Note that
the amplitude modulus $|A_i|$ is a direct measure of the mean square
deviation $\sigma_{\phi_i}$, introduced in Sect.~\ref{sec:num} to
quantitatively characterize the departure from full synchronization
in the numerical realization of our system. In fact, for harmonic
motion, $\sigma_{\phi_i}=a|A_i|/\sqrt{2}$. The phase $\varphi_i$ of
the complex amplitude $A_i=|A_i| \exp ({\rm i} \varphi_i)$ measures
the phase shift of the oscillations of $\psi_i(t)$ with respect to
the external perturbation.

\begin{figure}
\resizebox{\columnwidth}{!}{\includegraphics*{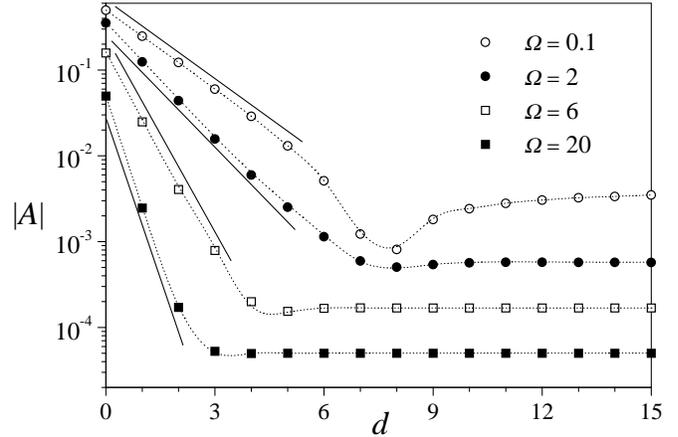} }
\caption{Amplitude moduli from the numerical solution of
Eq.~(\ref{matrix}), for the same system as in Fig.~\ref{f1}. Dotted
lines are spline approximations for $\sqrt{2}\sigma_{\phi}/a$, for
the values of the mean square deviation $\sigma_{\phi}$ presented in
Fig.~\ref{f1}. Full straight lines show the analytical prediction
for the slope at small distances, Eq.~(\ref{Aiap}).} \label{f6}
\end{figure}

Figure \ref{f6} shows the amplitude moduli $|A_i|$ obtained from the
numerical solution of Eq.~(\ref{matrix}), for a ensemble of $N=10^3$
oscillators with the same interaction network as in the results
presented in Sect.~\ref{sec:num}. The values of $|A_i|$ are averaged
over all the oscillators at a given distance $d$ from oscillator
$1$, and the subindex $i$ is accordingly dropped. Results are
presented for several values of the frequency $\Omega$. Dotted lines
are spline approximations of the numerical data for $\sigma_{\phi}$
already presented in Fig.~\ref{f1}, multiplied by a factor
$\sqrt{2}/a$. The agreement with the solution of Eq.~(\ref{matrix})
is excellent.

\begin{figure}
\resizebox{\columnwidth}{!}{\includegraphics*{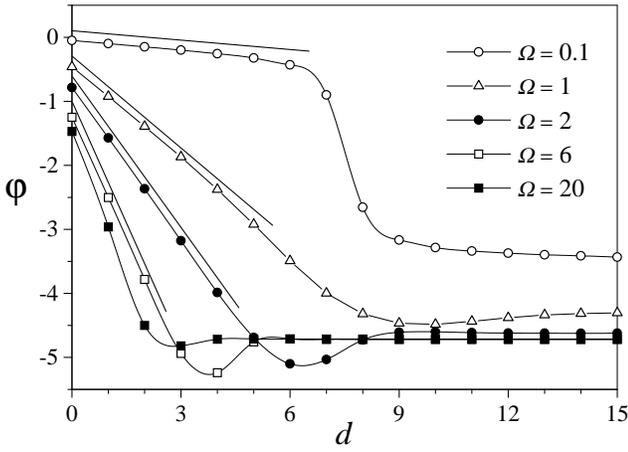} }
\caption{Phase shift as a function of distance, from the numerical
solution of Eq.~(\ref{matrix}) and for the same system as in
Fig.~\ref{f1}. Straight lines show the analytical prediction for the
slope at small distances, Eq.~(\ref{Aiap}).} \label{f7}
\end{figure}

The phase shifts $\varphi_i$ corresponding to the same solutions of
Eq.~(\ref{matrix}) are shown in Fig.~\ref{f7}, as a function of the
distance and for several values of the frequency $\Omega$. Data for
$\Omega=1$ are also shown. As it may be expected, phase shifts are
always negative, indicating a delay in the response of the system to
the external perturbation. Coinciding with the regime where the
amplitude moduli $|A_i|$ decrease exponentially, we find a zone where
phase shifts vary linearly with distance. Namely, the phase shift
between oscillators whose distances to oscillator $1$ differ by one
is a constant $\Delta \varphi$. This unitary phase shift, which gives
the slope of $\varphi$ as a function of the distance, depends on the
frequency $\Omega$. In this zone, individual phases vary with
distance as $\psi_i \propto \exp[{\rm i}(d_i \Delta \varphi+\Omega
t)]$. Thus, the perturbation propagates through the system at
constant speed $|\Omega / \Delta \varphi|$. For larger distances,
this linear regime breaks down, and the variation of $\varphi$ with
$d$ tends to be much less pronounced. In the transition between both
regimes, however, the phase shift varies rather abruptly for small
frequencies, while it develops a minimum for larger $\Omega$.

The results presented in Figs.~\ref{f6} and \ref{f7} where obtained
from the numerical solution of the small-amplitude limit equation
(\ref{matrix}). A more explicit solution can be obtained from a
suitable approximation of Eq.~(\ref{matrix}), taking into account
specific mathematical properties of the matrix $\cal L$, as we show
in the following section.

\subsection{Approximate solution}
\label{sec:anal2}

According to Eq.~(\ref{Lij}), the matrix $\cal L$ can be written as
\begin{equation}
{\cal L} = (z+{\rm i}\Omega) {\cal I}- \tilde {\cal J},
\end{equation}
where $\cal I$ is the $N\times N$ identity matrix, and the elements
of $\tilde {\cal J}$ are
\begin{equation} \label{Jt}
\tilde J_{ij}=J_{ij}-\frac{1}{N} \xi_j ,
\end{equation}
with $\xi_j=\sum_k J_{kj}$. Our approximation to the solution of
Eq.~(\ref{matrix}) is based on the following remarks.

(i) Due to the fact that $\sum_j J_{ij}=z$ for all $i$, the
eigenvalues of the adjacency matrix ${\cal J}$ are all less than or
equal to $z$ in modulus \cite{str}. Moreover, the eigenvalues of
$\tilde {\cal J}$ are the same as those of ${\cal J}$. In fact, if
${\bf v}=(v_1,v_2,\dots ,v_N)$ is an eigenvector of ${\cal J}$, then
$\tilde {\bf v}= (\tilde v_1,\tilde v_2,\dots ,\tilde v_N)$, with
$\tilde v_i=v_i-N^{-1}\sum_k v_k$, is an eigenvector of $\tilde {\cal
J}$ with the same eigenvalue. This implies that, for $\Omega\neq 0$,
the inverse of the matrix $\cal L$ can be expanded as
\begin{equation} \label{Linv}
{\cal L}^{-1}=\frac{[{\cal I}-(z+{\rm i}\Omega)^{-1} \tilde {\cal
J}]^{-1}}{(z+{\rm i}\Omega)} = \sum_{m=0}^\infty \frac{\tilde {\cal
J}^m}{(z+{\rm i}\Omega)^{m+1}},
\end{equation}
because all the eigenvalues of the matrix $(z+{\rm i}\Omega)^{-1}
\tilde {\cal J}$ are less than unity in modulus.

(ii) While $\sum_j J_{ij}=z$ for all $i$, $\sum_k J_{kj}=\xi_j $
varies with $j$. Note that $\xi_j$ is the number of links starting
at $j$, and thus gives the number of oscillators which are coupled to
oscillator $j$. For any realization of the interaction network,
however, the average value of $\xi_j$ over the whole ensemble is
always the same, $N^{-1}\sum_j \xi_j=N^{-1}\sum_{jk} J_{kj}=z$. This
suggests that, as an approximation to Eq.~(\ref{Jt}) avoiding the
explicit calculation of $\xi_j$, we can take $\tilde J_{ij}=
J_{ij}-z/N$. More generally, it is possible to show that this same
approximation yields, for the powers of $\tilde {\cal J}$,
\begin{equation} \label{Jtm}
\tilde J^{(m)}_{ij}=  J^{(m)}_{ij}-\frac{z^m}{N},
\end{equation}
where $J^{(m)}_{ij}$ and $\tilde  J^{(m)}_{ij}$ are elements of the
matrices ${\cal J}^m$ and $\tilde {\cal J}^m$, respectively.

Combining Eqs.~(\ref{Linv}) and (\ref{Jtm}), the approximate form of
matrix ${\cal L}^{-1}$ can be applied to the vector $\bf b$ in the
right-hand side of Eq.~(\ref{matrix}) to give the following
approximation for the amplitudes:
\begin{eqnarray}
A_i &=& \sum_{m=0}^\infty \frac{1}{(z+{\rm i}\Omega)^{m+1}} \left[
J^{(m)}_{i1} -\frac{z^m}{N} \right] \nonumber \\
&=& \sum_{m=0}^\infty \frac{J^{(m)}_{i1}}{(z+{\rm i}\Omega)^{m+1}}
+\frac{{\rm i}}{N\Omega }. \label{Ai}
\end{eqnarray}
In this approximation, the effect of the average deviation from full
synchronization --which, as we discussed in Sect.~\ref{sec:anal1},
can be interpreted as an external effective force acting over all the
oscillators with the same intensity-- is represented by the terms of
order $N^{-1}$.

It is interesting that the approximation (\ref{Ai}) involves
explicitly the matrix elements of the powers of ${\cal J}$. The
matrices ${\cal J}^m$ ($m=1,2,\dots$) bear information about the
metric structure of the interaction network. Specifically, the
element $J^{(m)}_{ij}$ equals the total number of directed paths of
length $m$ starting at node $j$ and ending at node $i$ \cite{gt}. In
other words, $J^{(m)}_{ij}$ gives the number of different ways of
reaching node $i$ from node $j$ in exactly $m$ steps along directed
links. Equation (\ref{Ai}) shows that the response of any individual
oscillator to the external perturbation, measured by the amplitude
$A_i$, is directly related to the number of paths though which the
perturbation signal can flow from oscillator $1$. Note that
$J^{(m)}_{i1}=0$ for $m<d_i$ and $J^{(m)}_{i1}=1$ for $m=d_i$. For
oscillator $i$, therefore, the first contribution to the sum in the
second line  of Eq.~(\ref{Ai}) comes from the term with $m=d_i$. For
$m>d_i$, $J^{(m)}_{i1}$ is different from zero if at less one path
of length $m$ starts at oscillator $1$ and ends at $i$.

For nodes at small distances from oscillator $1$, there is typically
only one path of length $d_i$ from $1$ to $i$. In fact, the
probability of having more than one path of short length between any
two oscillators is, at most, of order $z/N$ (which we assume to be a
small parameter, as in our numerical analysis). As a result, for
most oscillators at a small distance from oscillator $1$, we have
$J^{(d_i)}_{i1}=1$. Moreover, the total number of nodes with small
$d_i$, $n(d_i)\sim z^{d_i}$, is also small as compared with the
system size $N$.  This implies that the possibility that an
oscillator at a small distance $d_i$ is also connected by a path of
length slightly larger than $d_i$ can be neglected. Consequently,
for oscillators at small distances from the node at which the
perturbation is applied, the sum in the second line of
Eq.~(\ref{Ai}) is dominated by the term with $m=d_i$. This dominance
is enhanced for large $|z+{\rm i}\Omega|$, because successive terms
in the sum are weighted by increasing inverse powers of that number.
If the system is large enough we can drop the last term in the
second line of Eq.~(\ref{Ai}), and write $A_i \approx (z+{\rm
i}\Omega)^{-d_i-1}$, i. e.
\begin{equation} \label{Aiap}
A_i \approx (z^2+\Omega^2)^{-\frac{d_i+1}{2}} \exp \left[ -{\rm i}
(d_i+1) \tan^{-1} \frac{\Omega}{z} \right].
\end{equation}
Within this approximation, the small-distance exponential dependence
of the amplitude modulus, $|A_i| \approx (z^2+\Omega^2)^{
-\frac{d_i+1}{2}}$ is apparent. Straight lines in Fig.~(\ref{f6})
show the excellent agreement between the predicted slope of $|A_i|$
and our numerical results. Equation (\ref{Aiap}) also explains the
linear dependence of the phase shift $\varphi_i$ with the distance.
Specifically, it predicts a unitary phase shift $\Delta \varphi
=-\tan^{-1}(\Omega/z)$. Straight lines in Fig.~\ref{f7} stand for
this prediction.

Two effects contribute to break down the small-dist\-ance
approximation (\ref{Aiap}). First, as discussed above, we expect
that this approximation does not hold beyond distances where
$z^{d_i} \sim N$, i. e. $d_i \sim \log N / \log z$. At larger
distances, in fact, it is not true that only one path contributes to
the propagation of the perturbation signal from the node at which it
is applied. The second effect has to do with the relative magnitude
of the first non-zero term in the sum of the last line of
Eq.~(\ref{Ai}) and the term of order $N^{-1}$. For sufficiently
large distances, the latter cannot be neglected with respect to the
former. If $\Omega\ll z$, the two terms become comparable for
$z^{d_i+1} \sim N\Omega$, while if $z\ll \Omega$ they are similar for
$\Omega^{d_i} \sim N$. It turns out that, in both limits, this
second effect acts at distances smaller than the first one. This
does not imply, however, that the first effect plays no role in
determining the response of the system at large distances.

It can be shown that, for oscillators at large distances from
oscillator $1$, the matrix element in the sum of Eq.~(\ref{Ai}) can
be accurately approximated as $J^{(m)}_{i1}= J_0 z^m$ for all $m\ge
d_i$, while $J^{(m)}_{i1}= 0$ for  $m< d_i$. In fact, we can argue
that the number of paths of length $m$ ending at a given oscillator
$i$ scales as $z^m$ for large $m$, by noticing that this number is
$z$ times the number of paths of length $m-1$ ending at the
oscillators to which $i$ is coupled. The precise value of the
prefactor $J_0$ depends on $N$ and on the specific realization of
the network, but is independent of the distance $d_i$. For the
network corresponding to the numerical results presented above, for
instance, $J_0\approx 9.91 \times 10^{-4}$.

Replacing the {\it Ansatz} for $J^{(m)}_{i1}$ in Eq.~(\ref{Ai}) and
performing the summation, we get
\begin{equation} \label{Aiap1}
A_i \approx \frac{{\rm i}}{\Omega}\left[ -J_0
\left(1+\frac{\Omega^2}{z^2} \right)^{-\frac{d_i}{2} }\exp \left(
-{\rm i} d_i \tan^{-1} \frac{\Omega}{z} \right)+\frac{1}{N} \right] .
\end{equation}
This large-distance approximation for the complex amplitude $A_i$ is
more clearly analyzed for limit values of the frequency $\Omega$.
For small $\Omega$, specifically for $d_i \Omega/z \ll 1$, the
amplitude modulus is
\begin{equation}
|A_i| \approx \frac{1}{\Omega} \sqrt{\frac{d_i^2 \Omega^2}{z^2} J_0^2
+ \left(\frac{1}{N}-J_0 \right)^2}.
\end{equation}
As found in our numerical results, Figs.~(\ref{f1}) and (\ref{f6}),
the response grows with $d_i$ for large distances. Combined with the
decrease observed for small distances, this explains the existence
of an intermediate minimum in both $|A_i|$ and $\sigma_\phi$. The
amplitude phase
\begin{equation}
\varphi_i =-  \tan^{-1} \frac{z}{d_i\Omega}\left( \frac{1}{J_0
N}-1\right)
\end{equation}
exhibits a more complicated functional dependence with the distance.
In the opposite limit of large frequencies, the large-distance
approximation (\ref{Aiap1}) is dominated by the last term, $A_i
\approx {\rm i}/N\Omega$. We recall that this term stands for the
contribution of the average deviation from full synchronization to
the individual motion of oscillators. The amplitude modulus becomes
independent of the distance, as found in the results of
Fig.~\ref{f6}. Its phase is always $\pi/2$ --or, more generally,
$\pi/2+2k\pi$, with $k$ an integer. In the results for large
$\Omega$ shown in Fig.~\ref{f7}, we have $\varphi \approx \pi/2-2\pi
\approx -4.71.$

Coming back to the full form of our approximation for the amplitude,
Eq.~(\ref{Ai}), let us finally point out that, for sufficiently large
frequency $\Omega$ and irrespectively of the distance $d_i$, the
dominant terms are of order $\Omega^{-1}$. To this order, the sum
contributes its first term, $m=0$. Since ${\cal J}^0 ={\cal I}$, for
$\Omega\to \infty$  we get
\begin{equation}
A_i \approx \frac{\rm
i}{\Omega}\left(-\delta_{i1}+\frac{1}{N}\right).
\end{equation}
This result explains the decay as $\Omega^{-1}$ in the tails of the
resonance peaks, displayed in Fig.~\ref{f3}. It also shows that the
large-frequency response of oscillator $1$ is $N$ times larger than
that of any other oscillator, as illustrated in the same figure.
Moreover, we find that the phase shift of oscillator $1$ is
$\varphi_1=-\pi/2$ while any other oscillator is dephased by $\pi/2$
with respect to the external perturbation.

\section{Extensions}
\label{sec:ext}

\subsection{Other regular network structures}

It is important to remark that the analytical approach presented in
Sect.~\ref{sec:anal}, so far applied to random networks, is valid
for a large class of interaction patterns. In fact, the only
condition imposed on the adjacency matrix $\cal J$ to obtain the
results of Sect.~\ref{sec:anal1} is that $\sum_j J_{ij}=z$ for all
$i$, while the approximation of Sect.~\ref{sec:anal2} requires that
$\sum_i J_{ij}\approx z$ for all $j$. These conditions imply,
respectively, that each oscillator is coupled to exactly $z$
neighbours and that, in turn, the number of oscillators coupled to
each oscillator is also approximately constant. Under such
conditions, our approach can be used to evaluate the response of an
ensemble with any interaction pattern. In this section, we illustrate
this fact with a few cases that admit to be worked out explicitly.

Consider first a linear array of $N$ oscillators with periodic
boundary conditions, where each oscillator is coupled to its nearest
neighbour to the left ($z=1$). We assume that oscillators are
numbered from left to right in the natural order. For this directed
ring, we have $J_{ij}=1$ if $i=(j+1) \mbox{ mod } N$, and $J_{ij}=0$
otherwise. Consequently, $\sum_k J_{kj}=1$ for all $j$, and the
approximation of Sect.~\ref{sec:anal2} is exact. The relevant
elements of ${\cal J}^m$ are $J^{(m)}_{i1}=1$ if $m=d_i+kN$
($k=0,1,2,\dots$), and $J^{(m)}_{i1}=0$ otherwise. In fact, the only
paths that join oscillator $1$ with oscillator $i$,  at distance
$d_i=i-1$, are those of length $d_i$ plus an integer number $k$ of
turns around the ring. The calculation of the amplitudes yields
\begin{equation} \label{ring}
A_i=\frac{(1+{\rm i}\Omega)^{-d_i-1}}{1-(1+{\rm i}\Omega)^{-N}}+
\frac{\rm i}{N\Omega}.
\end{equation}
For $\Omega\ll 1$, but with $N\Omega \gg 1$, the amplitude at
essentially all distances ($d_i\lesssim N$) is dominated by the first
term of Eq.~\ref{ring}. In this regime, $|A_i|$ decreases
exponentially with $d_i$ and the phase shift $\varphi_i$ varies
linearly, with $\Delta \varphi=-\tan^{-1} \Omega$. In the opposite
limit of large frequencies, $\Omega\gg 1$, the amplitude decays as
$\Omega^{-1}$. For oscillator $1$ ($d_1=0$), we have $A_1 \approx
-{\rm i}/\Omega$, while for any other oscillator ($d_i>0$), $A_i
\approx {\rm i}/N\Omega$. We stress the qualitative similarity
between these results and those obtained for random networks.

Note that in the limit $N\to \infty$, the perturbation cannot attain
the oscillators to the left of oscillator $1$, an there is only one
path (of length $m=d_i$) for the signal to reach the oscillators to
its right. In this limit, $A_i= (1+{\rm i}\Omega)^{-d_i-1}$ at all
distances, and the regime of exponential decay and linear phase
shift extends over the whole system to the right of the node where
the perturbation is applied.

Consider now a linear array --which, for simplicity, we treat in the
limit $N\to \infty$-- where each oscillator is coupled to its two
first neighbours ($z=2$). Again, the approximation of
Sect.~\ref{sec:anal2} is exact. With this bidirectional coupling,
there are infinitely many directed paths joining any two nodes in
the network. A path of length $m$ between oscillator $1$ and
oscillator $i$ consists of $m_+$ steps to the right and $m_-$ steps
to the left, with $m_++m_-=m$ and $|m_+-m_-|=d_i$.  The number of
such paths is
\begin{equation} \label{binom}
J^{(m)}_{i1} = \left( \begin{array}{c} m \\ m_+ \end{array}\right)=
 \left( \begin{array}{c} m \\ \frac{m+d_i}{2} \end{array}\right)
\end{equation}
if $m$ and $d_i$ ($m \ge d_i$) are both even or odd, and $0$
otherwise. To perform the summation in Eq.~(\ref{Ai}), it is
convenient to write $m=d_i+2q$ and sum over $q=0,1,\dots$. This
yields
\begin{eqnarray}
A_i &=& \sum_{q=0}^\infty \frac{\left( \begin{array}{c} d_i+2q \\ q
\end{array}\right)}{(2+{\rm i}\Omega)^{d_i+2q+1}} \nonumber \\
&=& \frac{\ _2F_1 \left[ \frac{d_i+1}{2}, \frac{d_i}{2}+1;d_i+1;
\frac{4}{(2+{\rm i}\Omega)^2} \right]}{(2+{\rm i}\Omega)^{d_i+1}} ,
\label{2F1}
\end{eqnarray}
where $_2F_1(a,b;c;z)$ is the hypergeometric function \cite{abra}.
This exact result can be approximately analyzed for limiting values
of $\Omega$. For $\Omega\ll 1$, the amplitude modulus decreases
exponentially with the distance, as $ |A_i | \propto (1+
\sqrt{2\Omega} )^{-d_i/2}$, while the phase shift varies linearly
with $d_i$, with slope $\Delta \varphi = -\tan^{-1} (1+\sqrt{2/
\Omega})^{-1}$. Note that the exponential decrease of $|A_i|$ is much
slower than in the case of random networks with the same $z$. For
random networks, in fact, we have found that --in the limit of small
frequency and in the regime of exponential decay-- $|A_i| \sim
z^{-d_i-1}$. Here, on the other hand, the decay of $|A_i|$ becomes
increasingly slower as $\Omega \to 0$. This important difference is
a consequence of the fact that, in a random network and at small
distances, essentially only one path contributes to the propagation
of the perturbation signal towards each oscillator. For the
bidirectional linear array, in contrast, the total number of paths
$\sum_m J^{(m)}_{i1}$ grows exponentially with $d_i$. This growth
compensates partially the exponentially decreasing contributions
from successive path lengths [cf.~Eq.~(\ref{Ai})].

In the limit of large frequency, on the other hand, we find $A_i
\approx (2+{\rm i}\Omega)^{-d_i-1}$, i. e. $|A_i| \sim
\Omega^{-d_i-1}$. It is essential to this result, however, that
--assuming an infinitely large system-- we have neglected the term
of order $N^{-1}$ in Eq.~(\ref{Ai}). For any finite size, if $N$ is
fixed, the limit of large frequencies is dominated by this term, and
$|A_i| \sim \Omega^{-1}$.

Let us finally consider a interaction network with a hierarchical
structure, in the form of a directed tree where nodes are distributed
into successive layers. An oscillator at a given layer is coupled to
only one oscillator in the layer immediately above, so that $z=1$,
and the uppermost layer consists of a single oscillator, where the
external perturbation is applied. Thus, the perturbation propagates
downwards through the hierarchy. If layers are labeled in the
natural order starting by $l=0$ at the uppermost layer, the distance
$d_i$ of any oscillator to the node where the perturbation is
applied coincides with the label of its layer. Namely, oscillators
at level $l=1$ have $d_i=1$, at level $l=2$ they have $d_i=2$, and
so on. There is only one path, of length $d_i$, through which the
perturbation can reach oscillator $i$. Therefore, only one term
contributes to the sum in Eq.~(\ref{Ai}). Now, it is important to
note that the approximation (ii) introduced in Sect.~\ref{sec:anal2}
is no longer suitable. While each oscillator is coupled to only one
neighbour, the number of oscillators coupled to oscillator $j$,
$\sum_k J_{kj}$, is typically larger than one. In a regular
hierarchical structure, in fact, we fix $\sum_k J_{kj}=z'>1$ for all
$j$. In this situation, Eq.~(\ref{Ai}) changes in such a way that the
last term becomes multiplied by a factor $z'/z$. The amplitude is
\begin{equation}
A_i = (1+{\rm i}\Omega)^{-d_i-1}+\frac{z'}{z}\frac{{\rm i}}{N\Omega}
.
\end{equation}
Not unexpectedly, this result is similar to that for a
unidirectional linear array, Eq.~(\ref{ring}). In the first term,
which stands for the exponential-decay regime, the main difference
corresponds to a factor which, in the case of the array, takes into
account that the perturbation signal reaches an oscillator at each
turn around the array. In the second term, the two results differ
precisely in the factor $z'/z$. This effect enhances the response of
oscillators with $d_i>0$ at large frequencies, where $|A_i| \approx
z'/zN\Omega$.

\subsection{Non-regular random networks}

So far, our numerical and theoretical analyses have dealt with
regular networks, where all oscillators are coupled to exactly the
same number $z$ of neighbours. While the analytical approach cannot
be extended to the case of more general structures, it is worthwhile
to show that our results hold --at least, qualitatively-- for
non-regular random networks. For sufficiently large networks, with a
well defined average number of neighbours per site, it is in fact
expected that statistical quantities such as the mean square
deviation $\sigma_{\phi_i}$ are essentially not sensible to the
regularity of the interaction pattern.

We consider non-regular random networks of two types. In the first
type (random I) the number of neighbours  $z_i$ of each site $i$ is
chosen to be $1$, $2$ or $3$ with equal probability $1/3$. The
average number of neighbours is thus $\bar z= 2$, which makes it
possible to compare with our results for regular networks with
$z=2$. Once $z_i$ has been defined, the neighbours of site $i$ are
chosen at random from the whole system, avoiding self-connections
and multiple connections. In the second type of non-regular random
networks (random II), the number of neighbours of each site is drawn
from the discrete probability distribution $p(z) = 2^{-z}$
$(z=1,2,\dots) $, which also insures $\bar z= 2$ but with a much
wider dispersion. Our numerical calculations were run for a system
of $N=10^3$ oscillators, with a perturbation amplitude $a=10^{-3}$.

\begin{figure}
\resizebox{\columnwidth}{!}{\includegraphics*{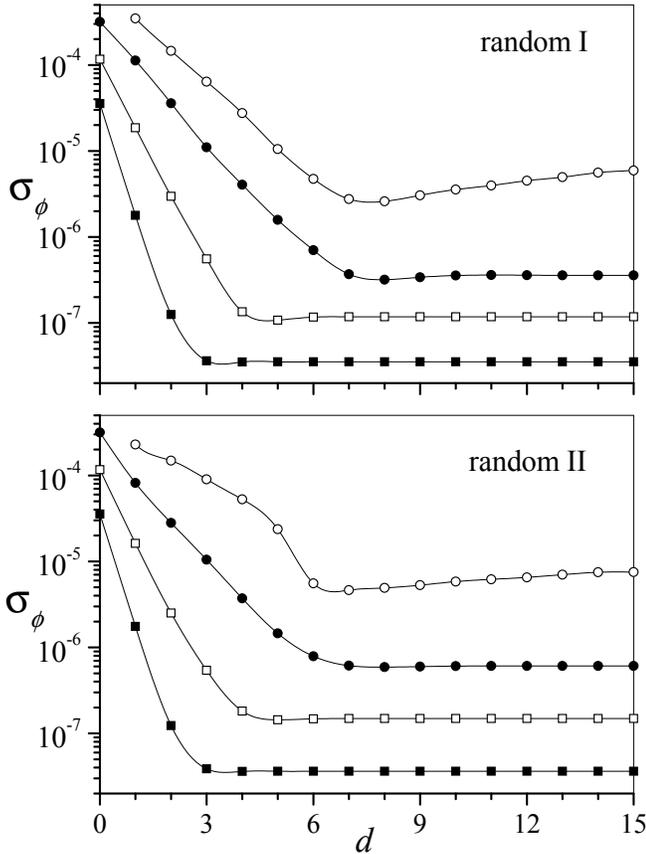}}
\caption{Mean square deviation from full synchronization in
non-regular random oscillator networks of the two types described in
the text (random I and II), for various values of the perturbation
frequency $\Omega$. Symbols are as in Fig.~\ref{f1}.} \label{fnew}
\end{figure}

Figure \ref{fnew} shows results for the mean square deviation from
full synchronization $\sigma_{\phi}$, with the two types of
non-regular random networks. For the sake of comparison with the
case of regular networks, the perturbation frequencies $\Omega$ are
the same as in Fig.~\ref{f1}. Moreover, specific realizations of the
networks with the same maximal distance to the perturbed oscillator,
$d_{\rm max}=15$, were selected. The vertical axes cover also the
same range. We verify at once that the main features in the
dependence of the mean square deviation on the distance found for
regular interaction patterns are also present in non-regular
networks. As it may have been expected, quantitative differences are
more important for small frequencies, i.e. near the resonance.
There, the oscillator network is more sensible to the perturbation
and, arguably, its detailed structure plays a more noticeable role
in determining its response. For larger frequencies, the values of
$\sigma_{\phi}$ become increasingly indistinguishable from those
obtained for regular networks.

\subsection{Chaotic oscillators}

An important question regarding the generality of the results
presented so far is whether they apply to ensembles of coupled
elements whose individual dynamics are not simple phase oscillations.
While we may argue that any cyclic behaviour, even in the presence
of external forces, can be approximately described by a periodic
phase oscillator \cite{kura}, the question remains open for chaotic
coupled dynamical systems. To address this problem we have
considered an ensemble of R\"ossler oscillators, described by the
equations
\begin{equation}
\begin{array}{ll}
\dot x_i &= -y_i-z_i+k\sum_jJ_{ij}(x_j-x_i) +a \delta_{i1}\sin
\Omega t \\ \\
\dot y_i &= x_i+0.2y_i +k\sum_jJ_{ij}(y_j-y_i)\\ \\
\dot z_i &= 0.2+z_i(x_i-c)+k\sum_jJ_{ij}(z_j-z_i) .\\
\end{array}
\end{equation}
The parameter $c$ controls the nature of the oscillations; for
$c=4.46$ they are chaotic \cite{MikhII}. As in Sects.~\ref{sec:num}
and \ref{sec:anal}, we choose the adjacency matrix such that $\sum_j
J_{ij}=z$ for all $i$. Each R\"ossler oscillator is thus coupled to
exactly $z$ neighbours. Chaotic systems can be fully synchronized if
the coupling intensity $k$ is larger than a certain threshold value,
related to the Lyapunov exponent of the individual dynamics
\cite{nos,str}. For the above value of $c$, a coupling intensity
$k=0.2$ insures that full synchronization is stable. In our ensemble
of R\"ossler oscillators, the external perturbation acts on just one
of the coordinates of oscillator $1$, namely, on $x_1(t)$.

\begin{figure}
\resizebox{\columnwidth}{!}{\includegraphics*{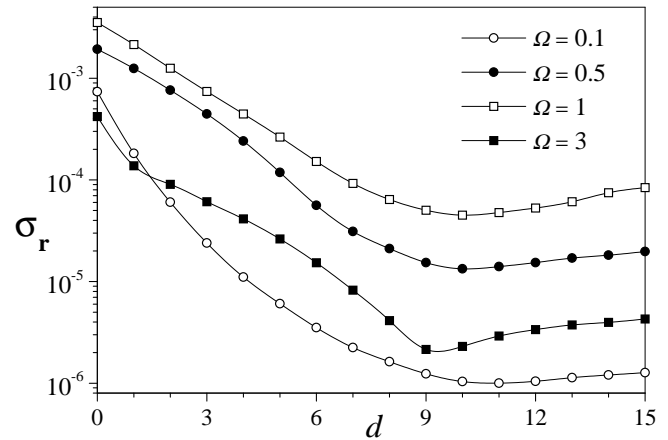}} \caption{Mean
square deviation from full synchronization as a function of the
distance for an ensemble of R\"ossler oscillators, for various values
of the frequency $\Omega$. Spline interpolations are shown as
curves.} \label{fr1}
\end{figure}

The numerical results presented below have been obtained for an
ensemble of $N=10^3$ R\"ossler oscillators, with $z=2$ and the
parameters quoted in the preceding paragraph. For the sake of
comparison, the interaction network is the same as in our study of
phase oscillators (Sects.~\ref{sec:num} and \ref{sec:anal}). The
amplitude of the external perturbation is $a=10^{-3}$. As a
characterization of the response of the system, we have used a
natural extension of the mean square deviation from full
synchronization defined in Eq.~(\ref{sigma}) for phase oscillators,
given by
\begin{equation}
\sigma_{{\bf r}_i} =  \left(\left\langle |{\bf r}_i - \bar{\bf r}|^2
\right\rangle \right)^{1/2},
\end{equation}
with ${\bf r}_i=(x_i,y_i,z_i)$ and $\bar {\bf r}=N^{-1} \sum_i {\bf r
}_i$. Time averages, indicated as $\langle \cdot \rangle$, are
performed over sufficiently long intervals, after transients have
been left to elapse. Figure \ref{fr1} shows numerical results for
the mean square deviation as a function of the distance, and for
various perturbation frequencies. Individual values of $\sigma_{{\bf
r}_i}$ for R\"ossler oscillators at a given distance from oscillator
$1$ have been averaged, and the index $i$ has been dropped
accordingly. We see that the overall behaviour of $\sigma_{\bf r}$
as a function of $d$  is similar to that found for ensembles of
phase oscillators (cf. Fig. \ref{f1}), with rapid decrease for small
distances and smooth growth for large distances. In all cases, the
two regimes are separated by a well defined minimum. The exponential
character of the decrease for small $d$ and the saturation of
$\sigma_{\bf r}$ for large $d$ are, however, much less clear than for
phase oscillators.

The results of Fig.~\ref{fr1} clearly show that the mean square
deviation varies non-monotonically with the perturbation frequency.
The dependence of $\sigma_{\bf r}$ with $\Omega$ is expected to
reveal the resonance nature of the system response. The natural
frequency $\omega$ of individual R\"ossler oscillators can be
defined, in the chaotic regime, in terms of the average period $T$
of the chaotic oscillations. Each period is determined, for
instance, as the time needed for an oscillator to cross the plane
$y=0$ in the subspace $x<0$. These times are then averaged over a
large number of oscillations, and the frequency is calculated as
$\omega=2\pi/T$. For $c=4.46$, we find $\omega \approx 1.08$.

\begin{figure}
\resizebox{\columnwidth}{!}{\includegraphics*{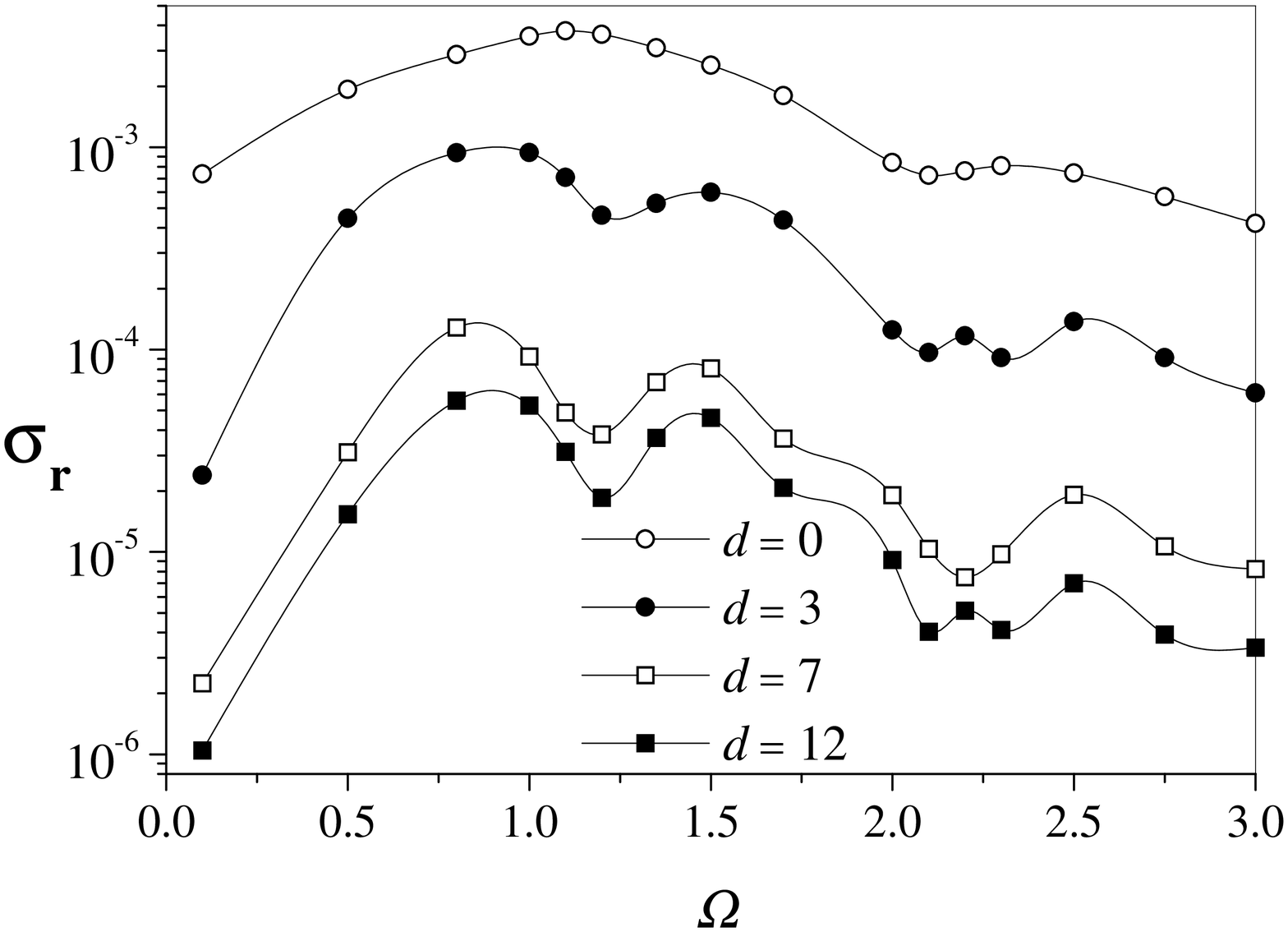}} \caption{Mean
square deviation from full synchronization as a function of
perturbation frequency for an ensemble of R\"ossler oscillators, at
various distances from the node where the perturbation is applied.
Spline interpolations are shown as curves.} \label{fr2}
\end{figure}

In Fig.~\ref{fr2} we present  numerical results for the mean square
deviation from full synchronization as a function of $\Omega$, for
several distances. For $d=0$, we find the expected resonance maximum
at $\Omega \approx \omega$. Interestingly enough, there is an
additional local maximum at $\Omega \approx 2\omega$, corresponding
to a harmonic resonance induced by non-linear effects. The presence
of this extra peak is consistent with the fact that higher-harmonic
components are very relevant contributions to the chaotic motion of
individual R\"ossler oscillators \cite{nonlin}. For larger
distances, the resonance maximum is replaced by a double peak, as we
have found for phase oscillators (Fig.~\ref{f2}), with a relative
minimum at $\Omega\approx \omega$ and two lateral maxima. At the site
of the harmonic resonance we find the same structure. In contrast
with the case of phase oscillators, however, the double peak persists
at large distances, with better defined maxima as $d$ grows.

We have also verified that, as for phase oscillators, the individual
motions of R\"ossler oscillators at the same distance from oscillator
$1$ are in-phase. These coherent dynamics give rise to a clustered
distribution in {\bf r}-space, and a snapshot of the ensemble in
that space produces a picture qualitatively very similar to
Fig.~\ref{f4}.

\section{Discussion and conclusion}

In this paper, we have studied the response of an ensemble of fully
synchronized oscillators to an external perturbation. The
perturbation is represented as an additional oscillator, evolving
autonomously with a fixed frequency. One of the oscillators of the
ensemble is coupled to this additional element. The perturbation
propagates through the system due to the coupling between
oscillators. In the absence of the external action, this interaction
sustains the state of full synchronization. The system can thus be
thought of as an active extended medium with a highly coherent rest
state --full synchronization-- whose response to the external
perturbation is driven by the collective dynamics of the interacting
oscillators. Such response, in fact, provides a characterization of
the collective dynamics.

Our study was mainly focused on ensembles of identical phase
oscillators. In its usual formulation, Kuramoto's model considers
global coupling, where interactions are identical for all oscillator
pairs. In this situation, all oscillators are mutually equivalent,
and the response to the external perturbation --other than on the
oscillator where the perturbation is applied-- is homogeneous over
the whole system.  Therefore, we have considered more complex
interaction patterns, introducing interaction networks which allow
for a non-trivial distribution of distances between oscillators.
Interactions were not necessarily bidirectional, so that coupling
was not always symmetric. To take advantage of certain analytical
results on the stability of full synchronization \cite{str}, we have
considered regular networks, were all oscillators are coupled to the
same number of neighbours. Numerical results obtained show, however,
that this choice does not represent a strong restriction on the
interaction network. The number of neighbours of each oscillator and
the frequency of the external perturbation are the main parameters
that control the response of the system.

For moderate values of its amplitude, the external perturbation
induces oscillations around the state of full synchronization. Our
main conclusion, first found by numerical means, is that the
response of each individual oscillator exhibits a clear dependence on
the distance from the node where the perturbation is applied. For
random interaction networks, this dependence shows two well defined
regimes. At small distances, the amplitude of the individual
oscillations decreases exponentially with the distance. Meanwhile,
the phase shift of these oscillations with respect to the external
perturbation, which measures the delay of the individual response,
varies linearly with the distance. In this regime, thus, the
perturbation propagates through the system at constant velocity, and
is progressively damped at a rate proportional to its own amplitude.
In other words, the system behaves as a linear dissipative medium.
For large distances, on the other hand, the individual response
saturates and the dependence of the amplitude and the phase shift
with the distance becomes much smoother. The relative extension of
the two regimes, as well as the rates of attenuation and dephasing
of the perturbation signal in the linear regime, depend on the
frequency of the external perturbation and on the number of
neighbours of each node.

The fact that the phase shift of individual oscillations with
respect to the external perturbation is defined by the distance to
the node where the perturbation is applied, implies that all the
oscillators at a given distance respond to the perturbation
coherently. Since the amplitudes of their oscillations are also
similar, the ``spatial'' distribution of the ensemble --i.e. the
distribution in the relevant one-particle state space-- becomes
clustered. Especially in the small-distance regime, where the
individual response  strongly depends on the distance, all the
oscillators at a given distance form a compact cluster with coherent
oscillatory motion. Thus, the external perturbation induces a
``spatial'' organization associated with the internal structure of
the interaction network.

The overall response of the system to the external perturbation is
maximal when the perturbation frequency is equal to the natural
frequency of the oscillators. This not unexpected resonance
phenomenon is revealed by the presence of a peak in the amplitude of
individual oscillations as a function of the perturbation frequency.
Far from the peak, the amplitude decreases as the inverse of this
frequency. For oscillators at intermediate distances, however, an
anomaly in the response appears. The resonance peak is replaced by a
double peak, with a minimum at the resonance frequency and two
lateral maxima. As we discuss below, this effect can be interpreted
as an interference phenomenon.

Our numerical results are well reproduced by an analytical approach
based on a linear approximation for small perturbation amplitudes.
This approach is able to discern between the roles of different
contributions to the response of the system. Two complementary
aspects are worth mentioning. First, we have found that the existence
of two regimes in the distance dependence is directly associated with
the distribution of the number of paths in the interaction network.
The perturbation signal reaches oscillators at small distances
essentially through only one path. The dissipative mechanisms
inherent to the dynamics of the oscillator ensemble progressively
attenuates the signal, which decays exponentially with the distance.
For oscillators at large distances, on the other hand, the number of
available paths grows exponentially, at a rate that --at least, for
small perturbation frequencies-- is similar to the rate of
exponential decay of the signal. These two partially compensating
effects determine that the variation of the response with the
distance is much smoother for large distances.

The second aspect has to do with the fact that the external
perturbation affects individual oscillators in two ways. Besides the
propagation of the signal through the network, which acts on the
oscillators with different intensities depending on their distance
to the node where the perturbation is applied, there is an overall
contribution originated by the average motion of the whole ensemble,
which affects all oscillators with the same intensity. This global
contribution is inversely proportional to the system size, and
therefore could be generally neglected for sufficiently large
systems. However, it does play an important role in determining the
system response at distances where the propagating signal has been
strongly damped. It also determines the response at high
perturbation frequencies, where the propagation mechanism is very
ineffective, especially, at large distances. Our analysis shows that
the overall contribution of the average motion of the ensemble and
the local contribution of the propagated signal have opposite signs.
In other words, their oscillation phases differ by $\pi$, in such a
way that, if their amplitudes are similar, a phenomenon of
destructive interference takes place. In our random networks this
happens, precisely, at the transition between the two regimes
discussed above, when the amplitude response has decreased by a
factor of about the inverse of the system size. The minimum in the
amplitude at those intermediate distances can thus be interpreted as
the result of the destructive interference between the two
contributions that affect individual motions.

Generally, we may expect that interference phenomena play an
important role in the dynamics of oscillator networks. This is due
to the fact that, as we have seen, the oscillatory signal changes
its phase as it propagates through the network. The signal can reach
a given oscillator through different paths, and thus with different
phases. The sum of all those contributions will depend not only on
their amplitudes but also on their relative dephasing, likely giving
rise to interference. From this perspective, the saturation of the
response at large distances, were contributions from many different
paths are acting, could be interpreted as a phenomenon of
constructive interference that breaks down the regime of exponential
decay.

The applicability of our analytical approach is not restricted to
random networks. We have shown how our main results extend to
regular and hierarchical arrays of phase oscillators. Numerical
calculations show, moreover, that the same results are qualitatively
reproduced in non-regular random interaction patterns. The most
important extension considered here, however, has consisted of
replacing phase oscillators by chaotic R\"ossler oscillators. The
response of ensembles of these chaotic elements to the external
perturbation was analyzed by numerical means. In spite of the
essential difference in the nature of the individual dynamics, we
have verified that the most important features found for phase
oscillators are qualitatively reproduced by R\"ossler oscillators.
Specifically, the existence of two regimes, depending on the
distance to the oscillator where the external perturbation is
applied, and the resonance nature of the response are also observed
for the chaotic elements. Non-linearities inherent to the chaotic
dynamics contribute extra effects, such as higher-harmonic
resonances.

As a concluding remark, let us stress that our analysis establishes
a close connection between the collective dynamics of an ensemble of
coupled oscillators subject to an external action and the
interaction pattern underlying the ensemble. This connection provides
a method to indirectly infer the structure of  such interaction
pattern by studying the response of individual oscillators to an
external perturbation. Sampling the motion induced by a perturbation
applied at different nodes on various oscillators may be used as an
experimental tool to reconstruct the interaction network.

\end{document}